\documentclass[aps,prd,twocolumn,groupedaddress,showpacs]{revtex4}

\usepackage{graphics,epsfig}

\begin{document}

\title{Is timing noise important in the gravitational wave detection
of neutron stars?}

\author{D. I. Jones} 
\email{jones@gravity.psu.edu}
\affiliation{Center for Gravitational Wave Physics, The Pennsylvania
State University, State College, Pennsylvania, 16802-6300}

\date{\today}

\begin{abstract}
In this paper we ask whether the phenomenon of timing noise long known
in electromagnetic pulsar astronomy is likely to be important in
gravitational wave (GW) observations of spinning-down neutron stars.
We find that timing noise is strong enough to be of importance only in
the young pulsars, which must have larger triaxialities than theory
predicts for their GW emission to be detectable.  However, assuming
that their GW emission is detectable, we list the pulsars for which
timing noise is important, either because it is strong enough that its
neglect by the observer would render the source undetectable, or else
because it is a measurable feature of the GW signal.  We also find
that timing noise places a limit on the observation duration of a
coherent blind GW search, and suggest that hierarchical search
techniques might be able to cope with this problem.  Demonstration of
the presence or absence of timing noise in the GW channel would give a
new probe of neutron star physics.
\end{abstract}


\maketitle

\section{Introduction}
\label{sect:i}

Spinning triaxial neutron stars may provide a source of detectable
gravitational radiation for the new generation of gravitational wave
interferometers \cite{jone02}.  The known pulsar population will be
targeted in forthcoming GW searches, as their locations are known to
high accuracy and their rotational properties have been studied in
some detail, greatly aiding the search.  The GW amplitudes of these
sources are likely to be low, so that only by coherently accumulating
signal over long time intervals (months to years) is there any hope of
making a positive detection \cite{thor87}.  This accumulation can be
achieved by the process of \emph{matched filtering}, where the noisy
detector output is multiplied by a template waveform which remains in
phase with the GW signal to better than one radian over the entire
observation span, and the resulting product integrated.

The spin frequencies of almost all pulsars are observed to gradually
decrease, presumably because of the loss of energy caused by
electromagnetic and (hopefully) GW emission \cite{lg98}.  Apart from
occasional glitches, this frequency change is gradual, so that pulsar
physicists can model the rotation phase using only a few terms of a
Taylor series of the form
\begin{equation}
 \label{eq:taylor}
 \Phi = 2\pi \int f_0 + \dot f_0 t + \ddot f_0 t^2/2 
		       + ...\, dt.
 \end{equation}
However, accurate radio timing observations reveal a small
irregularity in this spin-down known as \emph{timing noise}, occurring
on long timescales, comparable with the likely GW observation
timescales \cite{cord93}.  This noisy behavior of the electromagnetic
signal may well be present in the GW signal too, as almost all parts
of a neutron star are believed to be coupled together strongly on a
timescale of seconds or less \cite{saul89}.

If present in the GW signal the timing noise would result in a phase
difference between the idealized Taylor expansion and the real noisy
GW signal which grows in time (see Equation \ref{eq:DeltaPhiGW}
below).  It follows that there would exist a timescale $T_{\rm
decoherence}$ at which a simple Taylor series of the above form will
drift out of phase with the template by one radian, leading to a
complete loss of signal-to-noise and preventing detection.  This could
be prevented by the GW observer going to the trouble of using the
electromagnetically detected phase to generate the GW template.  A
method for achieving this has been devised by Pitkin \& Woan for the
Crab pulsar \cite{pw03}.

However, the timing noise should not be regarded simply as a nuisance,
complicating the detection process.  As described in section
\ref{sect:tpsotn}, the strength of the timing noise can be used to
probe stellar structure, giving a unique insight into neutron star
physics.  The accuracy with which the phase of the GW signal can be
extracted from the noisy data stream is proportional to one over the
signal-to-noise ratio \cite{chl03}, and so the GW phase measurement
error decreases as $T^{-1/2}$, where $T$ is the duration of the
coherent observation.  It follows that there exists a timescale
$T_{\rm detectable}$ at which the strength of the timing noise in the
GW data stream can be measured.

With these remarks in mind, we will pose and answer the following
three questions in this paper:
\begin{itemize}
\item
For each known pulsar, on what timescale $T_{\rm decoherence}$ would
the timing noise, if present in its GW signal, cause a simple Taylor
series search template to drift out of phase with the actual signal,
leading to a complete loss of signal-to-noise?
\item
For each known pulsar, on what timescale $T_{\rm detectable}$ would
the strength of GW timing noise be measurable?
\item
Is timing noise likely to be important when performing blind searches,
i.e. GW searches for neutron stars \emph{not} currently observed as
pulsars?
\end{itemize}
In all cases we have chosen to restrict our attention to observation
durations of \emph{three} years or less.  This is partly because the
computational costs of long coherent observations are very large
\cite{bccs98}, and partly because the interferometers are likely to be
upgraded significantly over such a timescale, in which case analysis
of a smaller set of newer, higher quality data might be more
profitable.

The structure of this paper is as follows.  We begin by summarizing
the physical significance of timing noise in section
\ref{sect:tpsotn}.  In section \ref{sect:moc} we assemble formulae
necessary to estimate $T_{\rm decoherence}$ and $T_{\rm detectable}$.
In section \ref{sect:ttn} we describe how we estimate the strength of
timing noise in the known pulsar population.  In section
\ref{sect:gwa} we provide both upper bounds and somewhat more
realistic estimates of the GW amplitudes of these pulsars.  In section
\ref{sect:rtdt} we estimate the decoherence timescales of the pulsars.
In section \ref{sect:r} we estimate possible timescales for
detectability of timing noise in the GW signal.  In section
\ref{sect:bs} we briefly discuss the importance of timing noise for
blind searches.  We summarize our results in section \ref{sect:c}.

\section{The physical significance of timing noise}
\label{sect:tpsotn}

Theoretically, the origin of timing noise is not understood.  Eight
different models of timing noise were considered in \cite{cg81}, where
the statistical properties of the timing residuals of each model were
calculated and compared with electromagnetic observations.  Some of
the models performed better than others, but a definitive
identification of the basic mechanism at work proved impossible.

Nevertheless, it is instructive to consider three rather general
scenarios, without linking them to any particular model of timing
noise.  In the first the whole star, including the pulsation-producing
magnetosphere, rotates as a rigid body, so that the relative phase of
the GW and electromagnetic wave (EW) is constant in time.  In the
second, the timing residuals represent a purely magnetospheric
phenomenon, where the location of pulsation production, at a height of
several neutron star radii above the surface, wanders randomly in
longitude, without there being any corresponding variation in the
rotational phase of the star.  In the third scenario, the variation in
phase is ascribed to a weak random angular momentum exchange between
the part of the star tied to the EW emission and the part tied to the
GW emission.  Denoting the moment of inertias of these two parts by
$I_{\rm EW}$ and $I_{\rm GW}$, and the \emph{departure} of their
angular velocities from a smooth spindown law by $\Delta \Omega_{\rm
EW}$ and $\Delta \Omega_{\rm GW}$, conservation of angular momentum
then demands
\begin{equation}
I_{\rm EW} \Delta\dot\Omega_{\rm EW} + I_{\rm GW}
\Delta\dot\Omega_{\rm GW} = 0.
\end{equation}
Integrating twice and rearranging gives a relationship between the GW
and EW phase residuals
\begin{equation}
\label{eq:phases}
\Delta\Phi_{\rm GW} = 
-\frac{I_{\rm EW}}{I_{\rm GW}} \Delta\Phi_{\rm EW}.
\end{equation}
(As we are considering triaxial neutron stars, the GW emission is at
\emph{twice} the rotation frequency \cite{mtw73}, so that the quantity
$\Delta\Phi_{\rm GW}$ in Equation (\ref{eq:phases}) is strictly only
one-half of the GW phase residual).  In this scenario, the
\emph{anti-correlation} of the gravitational and electromagnetic phase
residuals would then allow us to probe the relative moments of inertia
of the two parts of the star, as well as telling us that they are
loosely coupled on the timescale of the phase wandering.  For
instance, if these two parts were the crust and fluid core, we would
expect $I_{\rm EW}/I_{\rm GW} \sim 10^{-2}$ \cite{lrp93}.

Theoretically, all parts of the star (apart from any superfluid pinned
to the crust) are expected to be coupled strongly on timescales of the
order of seconds \cite{saul89}, orders of magnitudes smaller than the
timescales of years on which the phase residuals vary.  It therefore
seems most likely that the first of the above three scenarios is
correct.  However, given the potential importance of this phase
information, the following strategy suggests itself: Wherever
possible, the GW analysis of a known pulsar should use the observed
pulsar phase residuals $\Delta \Phi_{\rm EM}$ to demodulate the GW
detector output prior to using a smooth spindown template of the form
of equation (\ref{eq:taylor}).  A way in which such an analysis could
be performed for the Crab pulsar has been discussed recently by Pitkin
\& Woan \cite{pw03}.  The demodulation itself should correct for GW
phase wandering of the form
\begin{equation}
\label{eq:defnofalpha}
\Delta\Phi_{\rm GW} = \alpha \Delta\Phi_{\rm EW},
\end{equation}
with a range of different $\alpha$ values used.  The three scenarios
above correspond to the cases $\alpha = 1, 0, -I_{\rm EW}/I_{\rm GW}$,
respectively.  Measurement of a value of $\alpha$ other than unity
would challenge the standard picture of neutron star structure.

Of course, the addition of an extra parameter to an already
computationally expensive search should only be done if the extra
parameter is likely to have a significant affect upon the
signal---either by making what would otherwise be an undetectable
signal detectable, or by significantly improving the signal-to-noise
of an already detectable one.  It is precisely these issues that we
address in this paper, to assess whether or not GW observers need go
to the trouble of allowing for timing noise in their analyzes.

\section{Method of calculation}
\label{sect:moc}

Timing noise can be characterized as a random walk in one or a
combination of the rotation phase, frequency or spindown rate
\cite{cord93}.  These three idealized behaviors are known as phase
noise, frequency noise and spindown noise.  Pulsar physicists quantify
the strength of timing noise by $<\!\Delta \Phi_{\rm EW}^2\!>^{1/2}$, the
root mean square phase residual separating the actual signal and the
best Taylor series approximant to it, typically containing terms up to
and including the first frequency derivative.  For the three idealized
forms of random walk this rms phase residual grows as:
\begin{equation}
\label{eq:DeltaPhiEW}
<\! \Delta \Phi_{\rm EW}^2\!>^{1/2} = k T^{n/2},
\end{equation}
where $n=1,3,5$ for phase noise, frequency noise and slow-down noise,
respectively.  The corresponding GW timing noise in our model is then:
\begin{equation}
\label{eq:DeltaPhiGW}
<\!\Delta \Phi_{\rm GW}^2\!>^{1/2} = |\alpha| k T^{n/2}.
\end{equation}

\subsection{Calculation of the decoherence timescale}

Taking our criterion for total decoherence to be \mbox{$<\!\Delta \Phi_{\rm
GW}^2\!>^{1/2} = 1$}\,radian, Equation (\ref{eq:DeltaPhiGW}) gives
\begin{equation}
\label{eq:Tdecoherence}
T_{\rm decoherence} = \left(\frac{1}{|\alpha| k}\right)^{2/n}.
\end{equation}

\subsection{Calculation of the detection timescale}

A constant frequency periodic GW source is described by 7 parameters:
two angles giving its direction in the sky, its frequency, the phase
at time $t=0$, two angles specifying the orientation of its spin axis,
and its amplitude \cite{bg96}. When searching for GWs from a known pulsar,
the first four of these are known to high accuracy, so only the last
three need be fit for.

When timing noise is present in the GW signal (at least) two possible
strategies present themselves.  Firstly, we could imagine taking the
total data stream and breaking it up into a small number of equal
duration blocks, sufficiently short that each block can be analyzed
without worrying about the effects of timing noise.  The timing noise
would then manifest itself as a smooth variation in the phase at $t=0$
from block to block.  This method has the advantage that we need make
no assumptions about the relative phase of the EW and GW timing
noises, i.e. does not assume a correlation of the form of Equation
(\ref{eq:defnofalpha}), but the disadvantage that if the number of
data blocks is large the signal-to-noise ratio for each will be small,
leading to inaccurate phase measurements.

We will pursue a second, somewhat simpler strategy in this paper.  We
will assume that the EW and GW timing noises are correlated as in
Equation (\ref{eq:defnofalpha}).  Then $\alpha$ must be added as an
eighth search parameter.  This method has the advantage that the
entire signal-to-noise of the observation can be brought to bear on
the analysis.  Of course, if the EW and GW timing noises are not
correlated we would not expect any value of $\alpha$ to significantly
maximize the signal-to-noise ratio, indicating that we must use the
first method described above.

We present below a simple estimate of how long an observation time is
required to measure $\alpha$, i.e. to obtain a value significantly
different from zero.

The error in measuring the phase of a periodic GW signal of
signal-to-noise ratio $\rho$ is given approximately by
\begin{equation}
\delta \Phi_{\rm GW} = \frac{N}{\rho},
\end{equation}
where $N$ is a small number which depends upon the number of
parameters to be fit for \cite{chl03}.  The signal-to-noise grows as
$T^{1/2}$; we can make this time dependence explicit by writing
\begin{equation}
\label{eq:deltaPhiGW}
\delta \Phi_{\rm GW} \approx \frac{N}{\rho_{1-\rm yr} T_{{\rm yrs}}^{1/2}},
\end{equation}
where $\rho_{1\rm -yr}$ is the signal-to-noise that would be attained
for a one year observation, and $T_{\rm yrs}$ is the total observation
duration in years.

We will take as our criterion for the detectability of timing noise in
the GW signal that the rms phase wandering is equal to the GW phase
measurement error, i.e.  $<\!\Delta \Phi_{\rm GW}^2\!>^{1/2} \approx
\delta \Phi_{\rm GW}$.  Combining Equations (\ref{eq:DeltaPhiGW}) and
(\ref{eq:deltaPhiGW}) gives the time $T_{\rm equal}$ at which this
occurs
\begin{equation}
\label{eq:Tequal}
T_{\rm equal, \, yrs} = 
\left(\frac{N}{|\alpha| k\rho_{\rm 1-yr}}\right)^{2/(n+1)}.
\end{equation}
Of course, it is also necessary that the GW signal be detectable.  Let
us set the minimum signal-to-noise ratio for detection to some value
$\rho_{\rm min}$.  Define this signal-to-noise to be attained for an
observation duration $T_{\rm GW}$.  Using the result that the
signal-to-noise grows as $T^{1/2}$ we have:
\begin{equation}
\label{eq:Tgw}
T_{\rm GW} = \left(\frac{\rho_{\rm min}}{\rho_{1-\rm yr}}\right)^2.
\end{equation}
The timing noise is only detectable when its magnitude exceeds the GW
phase measurement error (i.e. $T > T_{\rm equal}$) \emph{and} the GW
signal is detectable (i.e. $T > T_{\rm GW}$).  It therefore follows
that the presence of timing noise can be detected in the GW data
stream after an observation time $T_{\rm detectable}$ equal to the
\emph{maximum} of these two values:
\begin{equation}
\label{eq:Tdetectable}
T_{\rm detectable} = \max (T_{\rm equal}, T_{\rm GW}).
\end{equation}

We would like to apply the above formulae to the entire pulsar
population, estimating $T_{\rm detectable}$ and $T_{\rm decoherence}$
for each known pulsar.  In order to make use of the formulae, we need
to obtain estimates of the timing noise, parameterized by $k$ and $n$,
and also of the GW signal strength, parameterized by $\rho_{\rm
1-yr}$.

\section{The timing noise}
\label{sect:ttn}

Ideally we would use observationally derived timing noise parameters
$k$ and $n$ for the entire pulsar population.  Unfortunately, a
literature search shows that only a minority of pulsars have been
timed in sufficient detail to allow for estimation of the timing noise
behavior.  However, on the basis of the available data, Dewey \&
Cordes \cite{dc89} have obtained a fitting formula that allows
estimation of the strength of the timing noise of a pulsar in terms of
its period $P$ and period derivative $\dot P$.  We will therefore
pursue the following strategy: For pulsars whose timing noise strength
has been measured, we will use the observational data in our analysis.
For pulsars whose timing noise hasn't been measured (or at least
weren't provided by our literature search), we'll use the fitting
formula of Dewey \& Cordes.

\subsection{Pulsars of measured timing noise}
\label{sect:pomtn}

A number of authors have published rms phase residuals $<\! \Delta
\Phi_{\rm EW}^2\!>^{1/2}$ for pulsars, typically after having fit the
timing data to a Taylor expansion including the frequency and its
first time derivative, over an interval of several years.  The
parameter $k$ can then be derived using the published values of $<\!
\Delta \Phi_{\rm EW}^2\!>^{1/2}$, $T$ and $n$ in Equation
(\ref{eq:DeltaPhiGW}):
\begin{equation}
\label{eq:kobs}
k = \frac{<\!\Delta \Phi_{\rm EW}^2\!>^{1/2}}{T^{n/2}}.
\end{equation}
Unfortunately, in only a small subset of these observations was it
possible to identify the nature of the walk, i.e. fix the value of
$n$.  For the sake of definiteness, we will therefore present results
for $n=3$; we have repeated calculations for $n=1$ and $n=5$ and found
results which are not very different, with the $n=3$ results being
intermediate between the two other sets.

The results of out literature search for timing noise data are
summarized in Table \ref{table:refs}.  For reasons of convenience, we
only selected pulsars of spin frequencies greater than $5$\,Hz,
corresponding to GW frequencies greater than $10$\,Hz, as stars
spinning more slowly than this will certainly not be of GW interest.
The references of table \ref{table:refs} then provided a total of $53$
pulsars.  Note that the majority of the pulsars selected come from the
Parkes Multibeam Study \cite{metal01, metal02,ketal03}, which was
specifically designed to find young pulsars.

\begin{table}
\caption{Papers containing radio pulsar timing noise data.  The number
of pulsars provided with $f_{\rm GW} > 10$\,Hz or greater is indicated.}
\label{table:refs}
\begin{tabular}{|l|l|} 
\hline
Reference & Number of pulsars \\
\hline
Cordes \& Downs (1985), \cite{cd85}      & 1  \\
Cordes \& Helfand (1980), \cite{ch80}   & 2  \\
D'Alessandro et al.\ (1995), \cite{dmhd95} & 2  \\
Cordes et al.\ (1988), \cite{cdk88}        & 1  \\
Manchester et al.\ (2001), \cite{metal01}    & 10  \\
Morrisat et al.\ (2002), \cite{metal02}       & 14  \\
Kramer et al.\ (2003), \cite{ketal03}         & 23  \\
\hline
\end{tabular}
\end{table}

\subsection{Pulsars whose timing noise hasn't been measured}
\label{sect:pwtnhbm}

In producing their fitting formula for timing noise, Dewey \& Cordes
make use of the `activity parameter', defined as the logarithm of the
rms phase residual for the pulsar in question divided by that of the
Crab:
\begin{equation}
\label{eq:A}
A = \log\left(\frac{<\!\Delta \Phi_{\rm EW}^2\!>^{1/2}}
    {<\!\Delta \Phi_{\rm EW, \, Crab}^2\!>^{1/2}}\right).
\end{equation}
The fitting formula is
\begin{equation}
\label{eq:Afit}
A = -1.4\log P + 0.8\log\dot P_{\rm{-15}} - 3.31.
\end{equation}
where $P$ is the pulsar's spin period in seconds and $\dot P$ is the
dimensionless period derivative divided by $10^{-15}$.  The Crab is
observed to display frequency-type timing noise of the form:
\begin{equation}
<\!\Delta \Phi_{\rm EW, \, Crab}^2\!>^{1/2} =
 0.24 {\rm \, \,  radians \, \,} T_{\rm yrs}^{3/2}
\end{equation}
(see \cite{cord93}).  It follows that the activity parameter only
makes sense for (Crab-like) frequency-type noise, as $A$ would
otherwise be a function of time. We will therefore assume that all
pulsars of measured timing noise display frequency-type noise.  The
above three equations can then be combined to give an estimate of the
rms timing noise in any pulsar of known period and period derivative.
Using $<\!\Delta \Phi_{\rm EW}^2\!>^{1/2} / <\!\Delta \Phi_{\rm EW, \,
Crab}^2\!>^{1/2} = k/k_{\rm Crab}$ in Equation (\ref{eq:A}) then
allows calculation of $k$:
\begin{equation}
k  = k_{\rm Crab} 10^{A}.
\end{equation}
Application of this formula to pulsars of known timing noise shows
that there exists a scatter of about an order of magnitude in the
measured activity parameter values about the predicted ones
\cite{cord93}.  This uncertainty should be borne in mind when reading
the results sections of this paper---individual pulsars may be more or
less noisy than assumed.

We used the above prescription for the $1182$ pulsars listed in the
Australia Telescope National Facility database \cite{atnf} whose
periods, period derivatives and distances from Earth were known,
excluding the $53$ of measured timing noise described in section
\ref{sect:pomtn}, i.e. the fitting formula was used for $1182-53 =
1129$ pulsars.

\section{Gravitational wave amplitudes}
\label{sect:gwa}

\begin{figure}
\epsfig{file=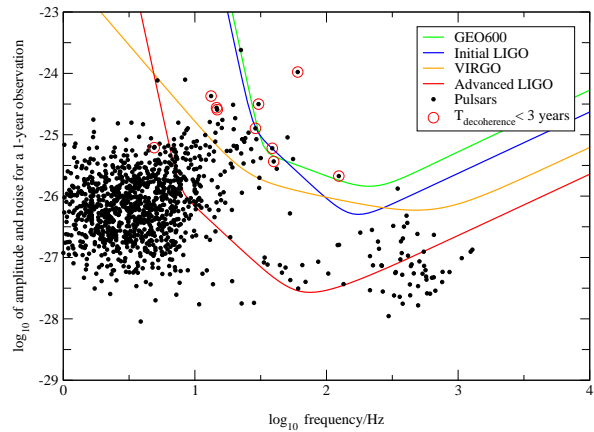,width=7cm,angle=-90}
\caption{Upper bounds on GW amplitudes assuming 100\% conversion of
spindown energy into GWs.  The noisecurves are for a one year
observation, and were produced using the fitting formulae of
\cite{os99}.  The ten red circles indicate those pulsars which have
$T_{\rm decoherence} < 10$\, years; see Section \ref{sect:rtdt}.
\label{fig:trifig}}
\end{figure}

We also need to estimate the GW amplitudes of the pulsar population,
giving the result in the form of the signal-to-noise for a one year
coherent integration.  The GW amplitude of a triaxial star a distance
$r$ from Earth is given by:
\begin{equation}
\label{eq:h}
h = \left(\frac{2}{15}\right)^{1/2} \frac{G}{c^4} \frac{8\Omega^2}{r} 
\Delta I,
\end{equation}
where the normalization comes from averaging over all possible spin
orientations of the source and $\Delta I$ is the difference between
the two principal moments of inertia in the plane orthogonal to the
spin \cite{thor87}.  This is more conveniently expressed as a
dimensionless number
\begin{equation}
\label{eq:defepsilon}
\epsilon = \frac{\Delta I}{I},
\end{equation}
where $I$ is the moment of inertia of the non-rotating star; we will
refer to $\epsilon$ as the \emph{ellipticity}.  Of course, the wave
amplitudes at Earth are unknown as $\epsilon$ is unknown.  An upper
bound is often placed on these quantities by assuming that all the the
spindown kinetic energy lost by the star is converted into GW energy,
at fixed moment of inertia.  In general the GW luminosity is given by
\begin{equation}
\label{eq:Edot}
\dot E = \frac{32G}{5c^5} \Omega^6 (\Delta I)^2,
\end{equation}
while the kinetic energy is $I \Omega^2/2$.  Combining these results
gives an upper bound on $h$ in terms of the pulsar's spin period,
period derivative and distance from Earth:
\begin{equation}
\label{eq:hspindown}
h_{\rm spindown} = 
\frac{2}{r} \left(\frac{G  I \dot P}{c^3 P}\right)^{1/2}
\end{equation}
with a corresponding ellipticity
\begin{equation}
\label{eq:espindown}
\epsilon_{\rm spindown} = 
\left[\frac{5c^5 \dot P P^3}{32 (2\pi)^4 G I}\right]^{1/2}.
\end{equation}
The results of such a calculation for all pulsars of known period,
period derivative and distance from Earth are shown in Figure
\ref{fig:trifig}.

\begin{figure}
\epsfig{file=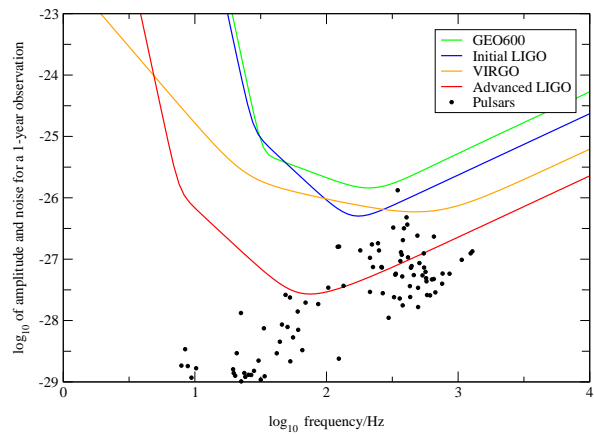,width=7cm,angle=-90}
\caption{Upper bounds on GW amplitudes imposing at cut-off of
$10^{-7}$ in $\epsilon$. \label{fig:epsilon_max}}
\end{figure}

The ellipticities required to produce this level of spindown are of
order $10^{-8}$ for the millisecond pulsars.  Detailed modeling
suggests that such an ellipticity could well be produced by strains in
the neutron star crust \cite{ucb00}.  However, for the younger
pulsars, which include all those which lie above the noisecurves of
the first generation detectors, the required ellipticities are much
larger, around $10^{-4}$.  Such a value is almost certainly
unphysically large, suggesting that GW play little role in the energy
budget of young neutron stars.  With this in mind, in Figure
\ref{fig:epsilon_max} we show another set of estimated wave
amplitudes, repeating the above calculation, but imposing a cut-off in
$\epsilon$ of $10^{-7}$, i.e. setting $\epsilon = \max (\epsilon_{\rm
spindown}, 10^{-7})$.
The value $\epsilon = 10^{-7}$ was chosen as it is at the upper end of
what is considered plausible on physical grounds \cite{ucb00}.  The
wave amplitudes calculated for the millisecond pulsars are mostly
unchanged, but the amplitudes for the young pulsars have been reduced
by several orders of magnitude (most have fallen off the bottom of the
figure).  In the following sections we will present results assuming
ellipticities ranging from $\epsilon_{\rm spindown}$ down to
$10^{-7}$.  (It is probable that in reality pulsar ellipticities are
even smaller than $10^{-7}$, but, as we shall see, we need not
consider smaller values for the purposes of examining timing noise).

\section{Results: The decoherence timescale}
\label{sect:rtdt}

We will now insert the timing noise strengths calculated in section
\ref{sect:ttn} into Equation \ref{eq:Tdecoherence} to obtain estimates
of the timescale on which timing noise would cause a simple Taylor
series to completely decohere from the GW signal, assuming that the
latter is perfectly locked in phase with the EW signal (i.e. $\alpha =
1$ in Equation [\ref{eq:defnofalpha}]).  These estimates are
completely independent of the GW strength---they depend only upon the
strength of the timing noise.

We find that 10 pulsars have decoherence timescales of less than three
years.  To gain some insight, we show their positions in the GW
amplitude--frequency plot in Figure \ref{fig:trifig} as open circles,
assuming 100\% conversion of spindown energy into GWs.  As expected,
most of these are young pulsars---only one has a GW frequency of less
than $10$\,Hz.  None of them are millisecond pulsars; the much weaker
timing noise of the latter class of star leads to their having
decoherence timescales many orders of magnitude longer that those of
the young pulsars.

However, these estimates are only of interest if the GW signal is
itself detectable.  To this end, we will present results first
assuming 100\% conversion of spindown energy into GWs, and then
assuming various bounds upon the allowed ellipticity.

\subsection{Decoherence timescale assuming 100\% conversion of 
            spindown energy into GWs; probably unphysical}
\label{sect:tdca}

As stated above, we find that 10 pulsars have decoherence timescales
of less than three years.  However, even when making the optimistic
assumption that $100\%$ of the spindown energy is being converted into
GW energy, some of these pulsars have GW detection timescales of more
than three years, and are therefore of no interest.  To allow for
this, the names, $T_{\rm decoherence}$ and $T_{\rm GW}$ values for the
pulsars which satisfy $T_{\rm decoherence} < 3$\,years \emph{and}
$T_{\rm GW}<3$\,years are given in Table \ref{table:Tdecoh_init}
assuming the Initial LIGO noisecurve, and the corresponding results
for Advanced LIGO are given in Table \ref{table:Tdecoh_adv}.

\begin{table}
\caption{Pulsars with $T_{\rm decoherence} < 3$\,years \emph{and}
$T_{\rm GW} < 3$\,years, assuming the Initial LIGO noisecurve,
$100\%$ conversion of spindown energy to GW energy, and $\alpha=1$.}
\label{table:Tdecoh_init}
\begin{tabular}{|l|l|l|} 
\hline
Name &  $T_{\rm decoherence}/$\,years &  $T_{\rm GW}/$\,years\\
\hline
J0205+6449 & 2.0e+00 & 2.8e+00\\ 
J0537-6910 & 1.1e+00 & 2.3e+00\\ 
B0531+21 & 2.6e+00 & 1.3e-02\\ 
\hline
\end{tabular}
\end{table}

\begin{table}
\caption{Pulsars with $T_{\rm decoherence} < 3$\,years \emph{and}
$T_{\rm GW} < 3$\,years, assuming the Advanced LIGO noisecurve,
$100\%$ conversion of spindown energy to GW energy, and $\alpha=1$}
\label{table:Tdecoh_adv}
\begin{tabular}{|l|l|l|} 
\hline
Name &  $T_{\rm decoherence}/$\,years &  $T_{\rm GW}/$\,years\\
\hline
J0205+6449 & 2.0e+00 & 1.4e-04\\ 
B0531+21 & 2.6e+00 & 1.9e-06\\
J0537-6910 & 1.1e+00 & 6.2e-03\\ 
B0540-69 & 9.5e-01 & 4.1e-03\\ 
J1124-5916 & 1.9e+00 & 3.7e-03\\ 
B1509-58 & 1.4e+00 & 2.0e-03\\ 
J1617-5055 & 2.5e+00 & 1.1e-03\\ 
J1930+1852 & 1.9e+00 & 3.3e-03\\ 
J2229+6114 & 2.6e+00 & 1.7e-03\\  
\hline
\end{tabular}
\end{table}

Tables \ref{table:Tdecoh_init} and \ref{table:Tdecoh_adv} represent
the maximal set of pulsars for which timing noise could prevent
detection.  In fact, detection will be prevented only if $T_{\rm
decoherence} < T_{\rm GW}$.  For the $100\%$ conversion of spindown
energy to GW energy assumed in Tables \ref{table:Tdecoh_init} and
\ref{table:Tdecoh_adv}, this is the case only for the Initial LIGO
observation of pulsars J0205+6449 and J0537-6910.

Note that only one pulsar has a predicted decoherence timescale of
less than one year, namely pulsar B0540-69, which has $T_{\rm
decoherence} = 0.95$\,years.  We therefore see that, regardless of the
GW signal strength, timing noise can only ruin a Taylor-series based
GW detection for timescales of order a year or longer.

Motivated by the discussion of section \ref{sect:tpsotn}, if we set
$|\alpha| = 10^{-2}$ in Equation (\ref{eq:Tdecoherence}) then, for
frequency type noise, the decoherence timescales will be greater than
those of Table \ref{table:Tdecoh_adv} by a factor of $100^{2/3}
\approx 21.5$; in this case there are no known pulsars with $T_{\rm
decoherence} < 3$\,years.

\subsection{Decoherence timescale assuming various upper bo\-unds 
            on the triaxiality}
\label{sect:tdcb}

Putting an upper bound on the ellipticity will have no effect on the
decoherence timescales but will increase the GW detection timescales,
so that for each pulsar in the Tables \ref{table:Tdecoh_init} and
\ref{table:Tdecoh_adv}, for some assumed $\epsilon_{\rm max}$, the
condition $T_{\rm decoherence} < T_{\rm GW}$ would be satisfied and
timing noise would prevent GW detection.  For instance, putting
$\epsilon_{\rm max} = 6 \times 10^{-7}$ the Advanced LIGO observation
of the Crab satisfies the condition.  However, repeating the analysis
for the more physically plausible case where $\epsilon_{\rm max} =
10^{-7}$, we find that the GW detection timescales of all pulsars in
the table exceed 3 years, even for the Advanced LIGO noisecurve.

To sum up, timing noise could render as many as 3 pulsars undetectable
using simple Taylor series for Initial LIGO, and as many as 9 for
Advanced LIGO, but GW emission from these pulsars is only detectable
if they have ellipticities in excess of those expected theoretically.

\section{Results: The detection timescale}
\label{sect:r}

We will now combine the timing noise estimates of section
\ref{sect:ttn} with the GW amplitudes of section \ref{sect:gwa} to
calculate the $T_{\rm detectable}$ values for all known pulsars.  We
will present results assuming both the Initial and the Advanced LIGO
noisecurves.  In the results that follow we have set the parameter
$N=3$ in Equation \ref{eq:deltaPhiGW}.  Cutler et al.\ (2003)
\cite{chl03} have shown that for LISA observations at least, $N < 1.5$
for $40\%$ of the possible spin orientations of the source, so our
setting $N$ to twice this value is surely a conservative estimate of
the phase error.  Also, we have set $\rho_{\rm min} = 5$ in Equation
(\ref{eq:Tgw}); for the directed searches considered here this will
give an acceptably low false alarm rate (Reference \cite{bccs98},
Eqn. 1.4).

Again, we will divide our results into two parts.  In the first part
we will assume complete conversion of spindown energy to GW energy.
For all but the millisecond pulsars these wave amplitudes are probably
unrealistically large, so that this calculation gives a safe
\emph{lower} bound on the detectability timescale.  We will then
consider more realistic scenarios, where the assumed triaxiality
$\epsilon$ is limited to some maximum value.

\subsection{Detection timescale assuming 100\% conversion of spindown energy
            into GWs; probably unphysical}
\label{sect:r1}

We will begin by assuming the the GW and EW signals are perfectly in
phase (i.e. $\alpha = 1$ in Equation [\ref{eq:defnofalpha}]).  In the
case of Initial LIGO, the (measured) timing noise and (estimated) GW
amplitude of the Crab are both large, so that its timing noise
detectability time is very short, just $0.54$\,years $\approx
200$\,days.  However, the significantly lower estimated GW amplitudes
of the other pulsars (see Figure \ref{fig:trifig}) leads to only two
having $T_{\rm detectable}$ values of less than $3$\,years.  The names
of all three pulsars with $T_{\rm detectable} < 3$\,years are
collected in Table \ref{table:T_detectable_init_ligo}
\begin{table}
\caption{Pulsars with $T_{\rm detectable} < 3$\,years, assuming $100\%$
conversion of spindown energy into GW energy, the Initial LIGO
noisecurve, and $\alpha = 1$.}
\label{table:T_detectable_init_ligo}
\begin{tabular}{|l|l|} 
\hline
Name &  $T_{\rm detectable}/$\,years \\
\hline
J0205+6449 & 2.8e+00\\ 
B0531+21 & 5.4e-01\\
J0537-6910 & 2.3e+00\\  
\hline
\end{tabular}
\end{table}

In the case of Advanced LIGO a total of $46$ pulsars have $T_{\rm
detectable} < 3$\,years; see Table \ref{table:T_detectable_adv_ligo}.
The detectability timescale for the Crab is very short, just $0.058$\,years.
\begin{table}
\caption{Pulsars with $T_{\rm detectable} < 3$\,years, assuming
$100\%$ conversion of spindown energy into GW energy, the
Advanced LIGO noisecurve, and $\alpha=1$.}
\label{table:T_detectable_adv_ligo}
\begin{tabular}{|l|l||l|l|} 
\hline
Name &  $T_{\rm detectable}/$\,yrs & Name &  $T_{\rm detectable}/$\,yrs \\
\hline
B0114+58 & 2.6e+00    & J1531-5610 & 2.0e+00\\
J0205+6449 & 1.4e-01  & B1610-50 & 1.4e+00\\ 
B0531+21 & 5.8e-02    & J1617-5055 & 2.8e-01\\ 
J0537-6910 & 2.3e-01  & J1637-4642 & 2.2e+00\\ 
B0540-69 & 1.9e-01    & J1702-4310 & 2.2e+00\\
J0633+1746 & 2.6e+00  & B1706-44 & 4.0e-01\\ 
B0833-45 & 2.4e-01    & B1727-33 & 1.2e+00\\ 
J0834-4159 & 2.8e+00  & J1740+1000 & 2.0e+00\\ 
J0855-4644 & 2.6e+00  & J1747-2958 & 5.7e-01\\ 
B0906-49 & 2.7e+00    & B1757-24 & 7.6e-01\\
J0940-5428 & 9.8e-01  & B1800-21 & 7.8e-01\\
J1015-5719 & 2.4e+00  & J1809-1917 & 2.8e+00\\ 
J1016-5857 & 1.1e+00  & B1823-13 & 6.8e-01\\
B1046-58 & 7.3e-01    & J1828-1101 & 1.9e+00\\
J1105-6107 & 9.9e-01  & B1830-08 & 2.2e+00\\ 
J1112-6103 & 1.6e+00  & J1837-0604 & 7.6e-01\\ 
J1124-5916 & 3.1e-01  & B1853+01 & 2.5e+00\\
B1259-63 & 1.7e+00    & J1913+1011 & 9.7e-01\\ 
B1338-62 & 1.6e+00    & J1930+1852 & 3.0e-01\\ 
J1420-6048 & 4.3e-01  & B1930+22 & 2.5e+00\\ 
J1509-5850 & 1.5e+00  & B1951+32 & 5.0e-01\\ 
B1509-58 & 2.1e-01    & J2021+3651 & 1.3e+00\\
J1524-5625 & 1.0e+00  & J2229+6114 & 3.2e-01\\ 
\hline
\end{tabular}
\end{table}

A summary of these results is provided in the first line of table
\ref{table:summary}, where the number of pulsars with timing noise
detectability timescales of less than three years are shown, for both
Initial LIGO and Advanced LIGO.

\begin{table}
\caption{A summary of our results for the detection timescale.  The
number of pulsars with $T_{\rm detectable} < 3$\,years, for both
Initial LIGO and Advanced LIGO with $\alpha = 1$.  The first line
assumes $100\%$ conversion of spindown energy into GW energy, while
the remaining lines impose a cut-off in ellipticity as indicated.}
\label{table:summary}
\begin{tabular}{|c|c|c|} 
\hline
& \multicolumn{2}{c|}{Number of pulsars} \\ 
$\epsilon_{\rm max}$ & Initial LIGO & Advanced LIGO  \\
\hline
No bound  & 3 & 46 \\
$10^{-2}$ & 3 & 46 \\
$10^{-3}$ & 2 & 38 \\
$10^{-4}$ & 2 & 18 \\
$10^{-5}$ & 0 & 6  \\
$10^{-6}$ & 0 & 1  \\
$10^{-7}$ & 0 & 0  \\
\hline
\end{tabular}
\end{table}

Setting $|\alpha| = 10^{-2}$ leaves no pulsars with detectable timing
noise using Initial LIGO, and only $7$ for Advanced LIGO.  These are
listed in Table \ref{table:Tdetectablesmall}.
\begin{table}
\caption{Pulsars with $T_{\rm detectable} < 3$\,years, assuming
$100\%$ conversion of spindown energy into GW energy, and the
Advanced LIGO noisecurve, and $|\alpha| = 10^{-2}$.}
\label{table:Tdetectablesmall}
\begin{tabular}{|l|l|} 
\hline
Name &  $T_{\rm detectable}/$\,years \\
\hline
J0205+6449 & 1.4e+00\\ 
B0531+21 & 5.8e-01\\
J0537-6910 & 2.3e+00\\ 
B0540-69 & 1.9e+00\\ 
B0833-45 & 1.7e+00\\ 
B1509-58 & 2.1e+00\\ 
J1617-5055 & 2.8e+00\\ 
\hline
\end{tabular}
\end{table}

\subsection{Detection timescale assuming various upper bounds on the 
            triaxiality}
\label{sect:r2}

The number of pulsars with timing noise detectability timescales less
than three years, for both Initial LIGO and Advanced LIGO, are shown
in Table \ref{table:summary}, for $\epsilon_{\rm max} = 10^{-2},
10^{-3}, 10^{-4}, 10^{-5}, 10^{-6}, 10^{-7}$.  For reasons of brevity
we have not tabulated the names of the pulsars falling in these
categories; in any case they are a subset of the pulsars of Tables
\ref{table:T_detectable_init_ligo} and
\ref{table:T_detectable_adv_ligo}.  In the case where $\epsilon_{\rm
max}$ takes on the huge value of = $10^{-2}$, the results are the same
as in section \ref{sect:r1}, as the triaxialities required to give
$100\%$ GW spindown are less than this.  As $\epsilon_{\rm max}$ is
increased, the number of pulsars with $T_{\rm detectable} < 3$\,years
decreases, falling to zero at $\epsilon_{\rm max} = 10^{-7}$.
Clearly, timing noise will only be detectable in the GW signal if the
ellipticities of neutron stars are $10^{-6}$ or larger, i.e. at least
an order of magnitude greater than theoretical modeling currently
suggests.

\section{Blind searches}
\label{sect:bs}

A `blind search' is a search for gravitational waves from a neutron
star not currently known as a pulsar.  The presence of timing noise in
the GW signal for such a star is more pernicious than in the case of a
directed search.  In the latter case the electromagnetic pulsar data
can be used to demodulate the timing noise from the data stream.  In
the case of a blind search this option is not available, so that the
timing noise will impose a maximum duration on any blind search beyond
which the smooth Taylor series approximation to the phase errs by a
significant amount (i.e. by about one radian).  

In practice, a GW astronomer would perform a blind search by dividing
the sky up into small patches, such that the Doppler shifts induced by
the Earth's spin and orbital motion are negligible over the patch
\cite{bccs98}.  Then, for each such patch, a search over a range of
spindown parameters $\dot P$ is made.  From the parametrization of
equation (\ref{eq:Afit}), for a given spin period $P$ the strength of
the timing noise is then determined, so that the decoherence timescale
can then be estimated from equation (\ref{eq:Tdecoherence}).  The
maximum search duration is then given by the parametrized formula:
\begin{equation}
\label{eq:TmaxPPdot}
\log T_{\rm max} = 2.62 + 0.93\log P - 0.53\log \dot P_{-15},
\end{equation}
where $T_{\rm max}$ is measured in years, $P$ in seconds and $\dot
P_{-15}$ is the dimensionless period derivative divided by $10^{-15}$.
This formula could be used by GW observers to limit the duration of a
single coherent search.

However, there exists another limit on the length of a coherent blind
search---namely that imposed by the requirement of doing the search in
real time \emph{with finite computational power}.  Brady et al.\
\cite{bccs98} have investigated the limits placed on the length of a
single non-hierarchical coherent search in some detail.  They found
that for a star with a GW frequency $\alt 200$\,Hz and spindown
timescale $f/\dot f \agt 1000$\,years, the search is limited to a
duration of $\alt 18$\,days.  Inserting these spindown parameters into
Equation (\ref{eq:TmaxPPdot}) gives a maximum search duration of
$2.3$\,years, $47$ times longer.  For a GW frequency of $\alt 1$\,kHz
and spindown timescale $\agt 40$\,years, Brady et al find a maximum
search duration of just $0.8$\,days; timing noise gives a duration of
$9.5$\,days, $12$ times longer.  It therefore appears that the
finiteness of computational resources will place a more stringent
limit on the duration of a single non-hierarchical coherent blind
search than that due to timing noise.

However, the high computational costs of such searches have motivated
GW astronomers to devise more computationally efficient hierarchical
search techniques.  In these, the full data set of duration $T_{\rm
obs}$ is split up into $N$ shorter pieces of duration $T_{\rm short} =
T_{\rm obs}/N$.  Each short piece is then coherently analysed by
matched filtering, and the separate results then combined
incoherently.  There are two main formulations that have been proposed
to achieve this: the `stack-slide' method \cite{bc00} and the Hough
transform \cite{sp99}.  These will certainly be more computationally
efficient than the single coherent analysis described above.  For
instance, Schutz \& Papa \cite{sp99} found that a data set of $T_{\rm
obs} = 10^7$\,seconds could be searched for GW signals of
signal-to-noise ratio $\approx 23$ by splitting the data into blocks
of duration $T_{\rm short} = 14$\,hours, using a $20$\,Gflop computer,
although this did not include a search over spin-down parameters.

If it should prove that these hierarchical techniques are efficient
enough that $T_{\rm obs}$ should exceed the timing noise decoherence
timescale, then the following strategy could be used, of a type first
suggested by Brady \& Creighton \cite{bc00} in the context of variable
accretion rate systems: $T_{\rm short}$ could be chosen to be
significantly smaller than $T_{\rm max}$ of Equation
(\ref{eq:TmaxPPdot}), so that the short coherent searches are not
significantly degraded by the timing noise.  The incoherent stage must
then be performed allowing for the many possible phase shifts that
timing noise could introduce.  We suspect that this last step will be
computationally expensive as the timing noise can map out many
different paths through phase space, but we won't attempt to quantify
this cost here.

\section{Conclusions} 
\label{sect:c}

In this paper we have addressed three issues.

First we asked for which of the known pulsars might timing noise
prevent a GW detection, if the experimenter assumes only smooth
Taylor-series spindown.  Our main conclusions were as follows:
\begin{itemize}
\item
If the GW timing noise is at the same level as the EW timing noise
(i.e. $\alpha = 1$ in Eqn. [\ref{eq:defnofalpha}]), then, assuming
$100\%$ conversion of spindown energy into GW energy, as many as $3$
pulsars may be rendered undetectable by Initial LIGO (Table
\ref{table:Tdecoh_init}), and as many as $9$ for Advanced LIGO (Table
\ref{table:Tdecoh_adv}).
\item
If the GW timing noise is weaker, at the level of $|\alpha| = 10^{-2}$,
no pulsars will be rendered undetectable by timing noise.
\item
Dropping the assumption of $100\%$ conversion of spindown energy into
GW energy and instead placing an upper bound on the ellipticities
leads to fewer stars for which decoherence may be an issue.  Setting
$\epsilon_{\rm max} = 10^{-7}$ or smaller we find that timing noise
will not prevent the detection of any pulsars.
\end{itemize}

Next we asked for which of the known pulsars might timing noise be
strong enough to be observed in the GW data stream.  Our main
conclusions were as follows:
\begin{itemize}
\item
If the GW timing noise is at the same level as the EW timing noise
(i.e. $\alpha = 1$), then, assuming $100\%$ conversion of spindown
energy into GW energy, timing noise will be detectable in $3$ pulsars
by Initial LIGO (Table \ref{table:T_detectable_init_ligo}), and in
$46$ pulsars for Advanced LIGO (Table
\ref{table:T_detectable_adv_ligo}).
\item
If the GW timing noise is weaker, at the level of $|\alpha| = 10^{-2}$,
timing noise will not be detectable in the GW signal of any pulsars
using Initial LIGO, and in only $7$ using Advanced LIGO (Table
\ref{table:Tdetectablesmall}).
\item
Dropping the assumption of $100\%$ conversion of spindown energy into
GW energy and instead placing an upper bound on the ellipticities
leads to fewer stars for which timing noise is detectable (Table
\ref{table:summary}).  Setting $\epsilon_{\rm max} = 10^{-7}$ or
smaller we find that timing noise is not detectable in the GW signal
of any pulsars.
\end{itemize}

Finally we asked how timing noise might affect blind GW searches,
i.e. searches for GW emitters not currently observed as pulsars.  Our
main conclusion was that:
\begin{itemize}
\item
Timing noise places a limit on the length of a data set that can be
coherently analyzed (Equation \ref{eq:TmaxPPdot}).  Proposed
hierarchical search techniques could perform searches over longer
durations by allowing for the possibility of (many different possible
realisations of) timing noise when the short coherent analyses are
incoherently combined.
\end{itemize}

Table \ref{table:Tdecoh_init} is probably of most interest to today's
GW data analysts---it lists the pulsars for which timing noise is most
important for the first generation interferometers.  For these three
pulsars, GW astronomers may wish to use electromagnetic timing
residuals to improve their ability to search for these stars.  A
method of doing so for one of these (the Crab) was described recently
by Pitkin \& Woan \cite{pw03}.

To sum up, timing noise may be an important feature of the GW signal
of some tens of young pulsars, but these stars must have very large
ellipticities in order for the GW emission to be strong enough to be
detectable.  If observed in the GW signal, timing noise would provide
a new insight into neutron star dynamics.

\begin{acknowledgments}

It is a pleasure to thank Graham Woan for useful discussions during
this investigation, and the anonymous referee for pointing out an
omission.  The Center for Gravitational Wave Physics is supported by
the National Science Foundation under cooperative agreement PHY
01-14375.
\end{acknowledgments}

\bibliography{tn}

\end{document}